\documentclass[%
 reprint,
superscriptaddress,
 amsmath,amssymb,
 aps,
]{revtex4-2}

\usepackage{graphicx, subfigure, float}
\usepackage{dcolumn}
\usepackage{bm}
\usepackage{hyperref}
\usepackage{physics}
\usepackage{amsthm}
\usepackage{bbold}
\usepackage{thm-restate}

\newtheorem{res}{Result}
\newtheorem{defi}{Definition}

\begin{document}

\title{Interplays between classical and quantum entanglement-assisted communication scenarios}

\author{Carlos Vieira}
\email{carlos.humberto.vieira@outlook.com}
\affiliation{Departamento de Matem\'{a}tica Aplicada, Instituto de Matem\'{a}tica, Estat\'{i}stica e Computa\c{c}\~{a}o Cient\i fica, Universidade Estadual de Campinas, 13083-859, Campinas, S\~{a}o Paulo, Brazil}
\author{Carlos de Gois}%
 \email{carlos.bgois@uni-siegen.de}
\affiliation{%
Naturwissenschaftlich-Technische Fakultät, Universität Siegen, Walter-Flex-Stra\ss e 3, 57068 Siegen, Germany
}%

\author{Lucas Pollyceno}
 \email{lpolly@ifi.unicamp.br}
\affiliation{Instituto de F\'isica Gleb Wataghin, Universidade Estadual de Campinas, 130830-859, Campinas, Brazil
}%

\author{Rafael Rabelo}
\affiliation{Instituto de F\'isica Gleb Wataghin, Universidade Estadual de Campinas, 130830-859, Campinas, Brazil
}%

\date{\today}

\begin{abstract}
    Prepare-and-measure scenarios, in their many forms, can be seen as the basic building blocks of communication tasks. As such, they can be used to analyze a diversity of classical and quantum protocols --- of which dense coding and random access codes are key examples --- in a unified manner. In particular, the use of entanglement as a resource in prepare-and-measure scenarios have only recently started to be systematically investigated, and many crucial questions remain open. In this work, we explore such scenarios and provide answers to some seminal questions. More specifically, we show that, in scenarios where entanglement is a free resource, quantum messages are equivalent to classical ones with twice the capacity. We also prove that, in such scenarios, it is always advantageous for the parties to share entangled states of dimension greater than the transmitted message. Finally, we show that unsteerable states cannot provide advantages in classical communication tasks, thus proving that not all entangled states are useful resources in these scenarios.
\end{abstract}

\maketitle
\section{Introduction}
    Dense coding \cite{bennet1992communication}, random access codes (\textsc{rac}) \cite{wiesner1983conjugate,ambainis_dense_1999}, and quantum key distribution \cite{BB84} are outstanding communication protocols where quantum systems can be advantageous over their classical siblings. Behind the scenes, these communication tasks are particular instances of the prepare-and-measure (\textsc{pm}) scenario \cite{PhysRevLett.105.230501}. The building blocks of \textsc{pm} scenarios are a preparation device --- which produces and transmits a physical system --- and a measurement device, which receives the system and reads information out of it. Being the simplest correlation scenarios that presume communication, they are indispensable ingredients in quantum information processing protocols such as semi-device independent dimension certification \cite{PhysRevLett.105.230501, Wehner_Lower_2008, Bowles_Certifying_2014} and the analysis of quantum communication networks \cite{bowles2015testing,wang2019characterising}, while also providing means for self-testing states \cite{tavakoli2018self,miklin2020universal}, distributing quantum keys \cite{pawlowski2011semi}, certifying randomness \cite{passaro-randomness-2015} and playing an important part in the discussion of the informational principles of quantum theory \cite{pawlowski2009information,chaves2015information}.

    Such as many other correlation scenarios, \textsc{pm} quantum behaviors have advantages over classical behaviors in many communication tasks. To determine \emph{how} and \emph{which} quantum systems outperform their classical counterparts has been a cornerstone of quantum information in recent decades. As formalized in Sec. \ref{sec:preliminaries}, the possible behaviors in a \textsc{pm} scenario naturally depend on the available resources. For instance, by comparing quantum against classical communication aided by shared randomness, one may discuss nonclassicality. This is arguably the most studied instance of \textsc{pm} scenarios, but even so, there are overarching questions still unanswered. An example regards the connection between measurement incompatibility and nonclassicality. It has been previously shown that incompatible measurements are insufficient for the emergence of nonclassical behaviors in the \textsc{pm} scenario \cite{cgois_classicality_2021}, and in a particular case of random access codes it was proven to be necessary \cite{Carmeli_2020}. Although it is widely accepted that necessity holds for all \textsc{pm} scenarios, to the best of our knowledge there is still no detailed proof in the literature. One is discussed presented in Appendix \ref{sec:appendix-proof-incompatibility-pm}.

    More recently, a generalization to \emph{entanglement-assisted} prepare-and-measure (\textsc{ea-pm}) scenarios (Sec. \ref{sec:preliminaries-ea-pm-scenarios}) began to be systematically investigated \cite{Moreno_Semi_device_2021,tavakoli2021correlations,pauwels2021entanglement,Frenkel2022entanglement, Pauwels_adaptive_2022}. Similarly to standard \textsc{pm} scenarios, we may consider either quantum or classical communication, but this time the devices may include a pre-established correlation between them through a shared quantum state $\rho_{AB}$. In this case, entanglement enables higher performance in a handful of paradigmatic communication tasks, such as entanglement-assisted random access codes \cite{pawlowski2011semi} and, naturally, dense coding \cite{Moreno_Semi_device_2021}. Under the hood, these advantages mean that, depending on the allowed resources, the sets of behaviors can be significantly distinct. It is thus of paramount importance to understand how and why they differ.

    Section \ref{sec:results} is devoted to these questions. Our starting point is to compare the sets of entanglement-assisted behaviors with classical against quantum communication. In this regard, we derive a chain of inclusions between the sets of behaviors and show that, in the limit of arbitrary-dimensional entanglement, the correlations implied by quantum and classical communication are identical (Sec. \ref{sec:c-vs-q-comm}). Furthermore, we construct an instance of an ambiguous guessing game and show that, for any dimension of the communicated message, an extra pair of entangled qubits leads to better performance (Sec. \ref{sec:highdim}), and that some entangled states do not provide an advantage in classical communication tasks (Sec. \ref{sec:steering}), therefore solving two of the open questions posed in \cite{pauwels2021entanglement}.

\section{Preliminaries}
\label{sec:preliminaries}
\subsection{Prepare-and-measure scenarios}
\label{sec:preliminaries-pm-scenarios}

\begin{figure}[H]
        \centering
        \includegraphics[width=.66\columnwidth]{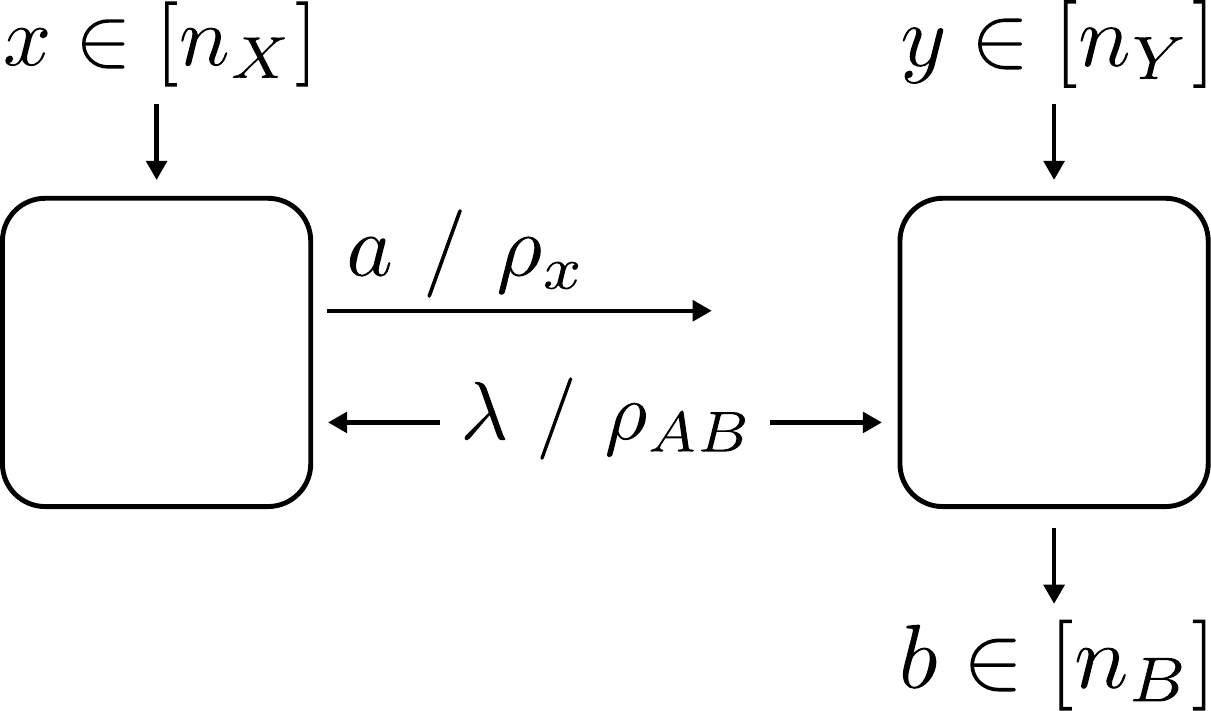}
        \caption{A preparation device, prompted by an input $x$, prepares and transmits either a classical or a quantum $d$-dimensional system to a measurement device. The latter is queried with $y$ and outputs $b$. The two devices may be correlated by a classical ($\lambda$) or quantum ($\rho_{AB}$) resource.}
        \label{fig:pm-scenarios}
    \end{figure}

    Prepare-and-measure scenarios can be interpreted as the encoding, transmission and decoding of information in a semi-device independent way. Let us say Alice receives a message $x \in [n_{X}] := \{0, \cdots, n_X - 1\}$, which she wants to communicate to Bob. She inputs her message to a preparation device that encodes it into a physical system, that is then transmitted to Bob's measurement device. This message, which may be either classical or quantum, is of dimension no greater than $d$. Bob then receives his own input $y\in [n_{Y}]$ which, provided to his measurement apparatus, results in an output $b \in [n_{B}]$ (Fig. \ref{fig:pm-scenarios}).
    
    To specify a prepare-and-measure scenario, we hence must provide the message dimension $d$ and a tuple of integer numbers $(n_{X}, n_{Y}, n_{B})$. Since the only observable quantities are $x$, $y$ and $b$, the result of this experiment is described by the family of conditional probability distributions $\{ p(b \vert x, y) \}_{b, x, y} $, the \emph{behavior} of the experiment, where $p(b \vert x, y)$ is the probability of Bob observing $b$ as output when Alice and Bob receive $x, y$ as their inputs, respectively.
    
    The central question in these scenarios is to compare the behaviors achievable given different resources. When a set of behaviors is a strict superset of another, we can say that the resources associated with the larger set provide advantage over some communication tasks. Apart from comparing the case of classical against quantum communication, we can furthermore distinguish the settings where the devices are independent, share classical correlations, or quantum correlations (which will be discussed further on). For independent devices, we will denote by $Q_{d}(n_{X}, n_{Y}, n_{B})$ and $C_{d}(n_{X}, n_{Y}, n_{B})$ the set of all behaviors $\{ p(b \vert x, y) \}_{b, x, y} $ obtained in a \textsc{pm} scenario $(n_{X}, n_{Y}, n_{B})$ having a qudit or a dit, respectively, as the transmitted system. When classical correlations (also called ``shared randomness'') are present, we denote the analogous sets by $\bar{Q}_{d}(n_{X}, n_{Y}, n_{B})$ and $\bar{C}_{d}(n_{X}, n_{Y}, n_{B})$. These sets are equivalent to the convex hulls of the corresponding $Q_d$ and $C_d$, respectively \cite{ambainis2008quantum, PhysRevA.86.042312}. 
    We will sometimes omit the labels $(n_{X}, n_{Y}, n_{B})$, and denote the classical and quantum sets only by $C_{d}$ and $Q_{d}$. In such cases, it should be clear that whenever we are comparing a set $C_{d}$ with a set $Q_{d'}$, the comparison is done with respect to the same set of labels $(n_{X}, n_{Y}, n_{B})$. More precisely:
    \begin{defi}[Quantum behaviors \cite{PhysRevLett.105.230501, PhysRevA.86.042312}]
            A behavior $\{ p(b \vert x, y) \}_{b, x, y} $ belongs to $Q_{d}(n_{X}, n_{Y}, n_{B})$ if there exists (i) a set of quantum states $\{\rho_x \}_{x\in [n_{X}]} \subset \mathcal{L}(\mathbb{C}^{d})$ and (ii) a set of measurements $\{\mathcal{M}_y\}_{y\in [n_{Y}]}$, where each $\mathcal{M}_y = \{ M_{b \vert y} \}_{b\in [n_{B}]}$ is a POVM, such that
            $$ p(b \vert x, y) = \Tr(\rho_x M_{b \vert y}),$$
        for all $x \in [n_{X}], y \in [n_{Y}]$ and $b \in [n_{B}]$. 
    \end{defi}

    \begin{defi}[Classical behaviors \cite{PhysRevLett.105.230501, PhysRevA.86.042312}]
        A behavior $\{ p(b \vert x, y) \}_{b, x, y} $ belongs to $C_{d}(n_{X}, n_{Y}, n_{B})$ if, for each $x\in [n_{X}]$, there exists (i) an encoding probability distribution $\{p(a|x)\}_{a \in [d]}$ for Alice, and (ii), for each $y\in [n_{Y}]$ and $a \in [d]$, a decoding probability distribution $\{p(b|y,a)\}_{b\in [n_{B}]}$ for Bob, such that
        \begin{equation}\label{eq:def-classical-behaviour}
            p(a \vert x, y) = \sum_{a=0}^{d-1}p(a|x)p(b|y,a),
        \end{equation}
        where the variable $a$ denotes the classical message sent by Alice. 
    \end{defi}
    
    \begin{figure*}\label{fig:eapm-behaviors}
        \centering
        \subfigure[Quantum communication]{\label{fig:ea-qpm}\includegraphics[width=.7\columnwidth]{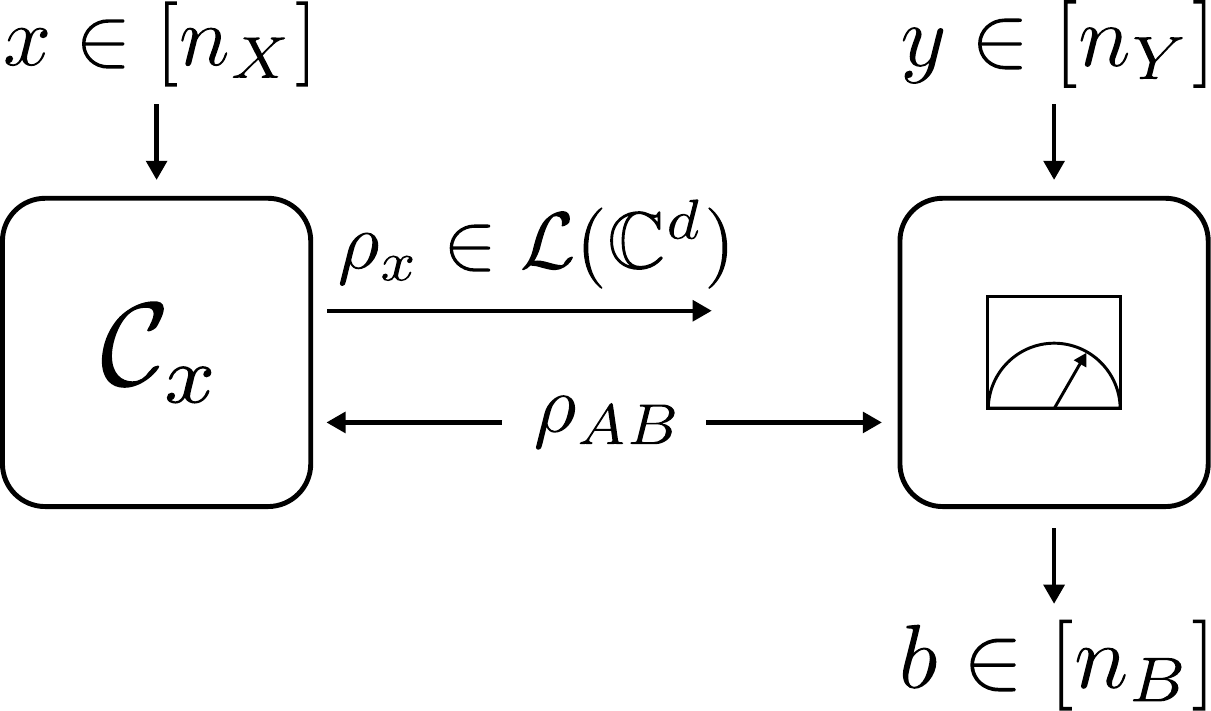}}\hspace{6em}
        \subfigure[Classical communication]{\label{fig:ea-cpm}\includegraphics[width=.7\columnwidth]{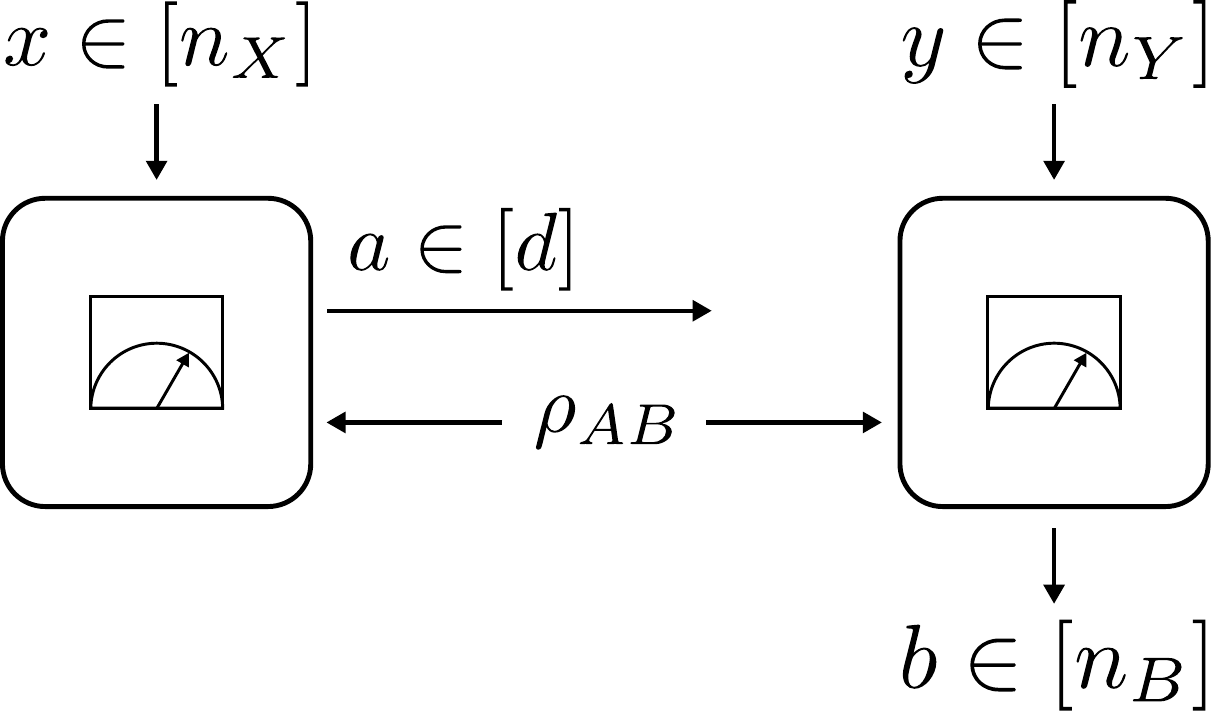}}
        \caption{In entanglement-assisted prepare-and-measure scenarios, a shared quantum resource $\rho_{AB}$ may be exploited to enhance the performance of quantum and classical communication tasks. When communication is quantum, a general behavior is described by a local, possibly dimension-reducing CPTP transformation ($\mathcal{C}_x$) on one share of the resource, which is then transmitted and measured. For classical communication, all behaviors can be described as Alice performing a local measurement on her share of $\rho_{AB}$, the output of which is then sent to the measurement device which, informed by the classical dit and Bob's choice of $y$, measures his share of $\rho_{AB}$.}
    \end{figure*}
    In the presence of shared randomness, the equations that define quantum and classical behaviors are given by
    \begin{equation}
        p(b \vert x, y) = \sum_{\lambda}\pi(\lambda)\Tr(\rho_{x}^{\lambda} M_{b \vert y}^{\lambda})
    \end{equation}
    and
    \begin{equation}\label{eq:behaviour_classical_SR}
         p(b \vert x, y) = \sum_{\lambda}\sum_{a=0}^{d-1}\pi(\lambda)p(a|x,\lambda)p(b|y,a,\lambda),
    \end{equation}
    respectively, where $\pi(\lambda)$ is a probability distribution. 
    
    It follows from the definitions that, regardless of the choices of $d, n_{X}, n_{Y}, n_{B}$, we always have $C_{d}(n_{X}, n_{Y}, n_{B}) \subseteq Q_{d}(n_{X}, n_{Y}, n_{B})$. Actually, for the classical case, we can consider that Alice and Bob hold quantum states and measurements that are all diagonal with respect to the same basis \cite{PhysRevA.86.042312, tavakoli_informationally_2022}. 
     Furthermore, in contrast with Holevo's bound, which states an equivalence between classical and quantum communication at the channel capacity level \cite{holevo1973bounds}, there is no equivalence between classical and quantum messages at the correlations level. That is, in general, $Q_{d}$ is not a subset of $C_{d}$. Indeed, one standard example is the scenario with $d = 2$ and $(n_X, n_Y, n_B) = (4, 2, 2)$, where quantum over classical advantage is obtained by means of random access coding \cite{ambainis_dense_1999}.
    
    Although, in general, $Q_{d}(n_{X}, n_{Y}, n_{B}) \nsubseteq C_{d}(n_{X}, n_{Y}, n_{B})$, there is a special class of \textsc{pm} scenarios where the inclusion is valid. Such a class is composed of all \textsc{pm} scenarios where $n_{Y} = 1$, that is, all \emph{communication channel} scenarios. Frenkel and Weiner showed that, in this case, all behaviors obtained with a quantum message can be simulated by sending a classical message of the same dimension \cite{frenkel-storage-2015}, i.e., that
    \begin{equation}\label{eq:result-Frenkel}
        \bar{C}_{d}(n_{X}, 1, n_{B}) = \bar{Q}_{d}(n_{X}, 1, n_{B}),
    \end{equation}
    for all choices of integers $d, n_{X}$ and $n_{Y}$. This observation --- which will be a central piece in much of our discussions, --- can be seen as a generalization of Holevo's bound, where in addition to the channel capacity of a qudit being the same as that of a dit, the set of behaviors are also equivalent.
    
    Interestingly, this result also implies that measurement incompatibility is a necessary condition for nonclassicality in prepare-and-measure scenarios. Here we take incompatibility as synonymous with non-joint measurability, and by ``nonclassicality'' we mean that quantum communication may provide some form of advantage over the classical case. Thus, only when Bob’s measurements are not jointly measurable, such advantage can be achieved. This fact has somewhat of a folklore status in the prepare-and-measure community but, 
    to the best of our knowledge, no formal proof of it has been published so far, except for special cases of \textsc{pm} scenarios, such as some random access coding tasks \cite{Carmeli_2020,Otfried_incompatible_2021}.
   For completeness, we provide a proof in Appendix \ref{sec:appendix-proof-incompatibility-pm}.
    
    Naturally, we may wonder whether incompatibility is also sufficient for nonclassicality in the prepare-and-measure scenario. However, this was proven to be false \cite{cgois_classicality_2021}. On the other hand, for \textsc{pm} scenarios where Alice receives quantum inputs, it was shown that measurement incompatibility is necessary and sufficient for nonclassicality \cite{Guerini_Distributed_2019}.

\subsection{Entanglement-assisted prepare-and-measure scenarios}
\label{sec:preliminaries-ea-pm-scenarios}

    A generalization of the previous scenario is found when Alice and Bob are allowed to share quantum correlations and employ them to increase performance in communication tasks. The use of entanglement as a resource in prepare-and-measure scenarios have only recently started to be systematically investigated, and many crucial questions remain open \cite{Moreno_Semi_device_2021,tavakoli2021correlations,pauwels2021entanglement, Frenkel2022entanglement, Pauwels_adaptive_2022}. In these entanglement-assisted prepare-and-measure (\textsc{ea-pm}) scenarios, the shared state $\rho_{AB}$ can be exploited to generate novel behaviors, which have been associated with quantum advantages in paradigmatic communication tasks, such as dense coding \cite{bennet1992communication} and random access codes \cite{PhysRevA.81.042326, Tavakoli_Spatial_2016}.
    
    When, in addition to a $D$-dimensional resource $\rho_{AB}$, communication is a qu$d$it system, any strategy in \textsc{ea-pm} can be described as follows. For each input $x$, Alice applies a CPTP map $\mathcal{C}_{x}: \mathcal{L}(\mathbb{C}^D) \to \mathcal{L}(\mathbb{C}^{d})$ to her share of the state $\rho_{AB}$. Alice then sends her transformed state to Bob, who performs one of the POVMs $\mathcal{M}_y = \{M_{b|y}\}_{b \in n_{B}}$ on the state $(\mathcal{C}_x \otimes \mathcal{I})(\rho_{AB})$ in his possession, where $\mathcal{I}$ is the identity channel. Bob's output, $b$, is the result of his measurement. More precisely, we define the set of behaviors $Q_{d}^{D}(n_{X}, n_{Y}, n_{B})$ as \cite{tavakoli2021correlations,pauwels2021entanglement} (see also Fig. \ref{fig:ea-qpm}):
    \begin{defi}[Entanglement-assisted quantum behaviors]\label{def: EA-pm-quantico}
        A behavior $\{ p(b \vert x, y) \}_{b, x, y} $ belongs to $Q_{d}^{D}(n_{X}, n_{Y}, n_{B})$ if there exists (i) a state $\rho_{AB} \in \mathcal{L}(\mathbb{C}^{D} \otimes \mathbb{C}^{D})$; (ii) a set of CPTP maps $\{\mathcal{C}_x \}_{x\in [n_{X}]}$, with $\mathcal{C}_x: \mathcal{L}(\mathbb{C}^{D}) \to \mathcal{L}(\mathbb{C}^{d})$; and (iii) a set of quantum measurements $\{\mathcal{M}_y\}_{y \in [n_Y]}$, where each $\mathcal{M}_y = \{ M_{b \vert y} \}_{b \in n_B}$ is a POVM, such that
        \begin{equation}\label{Eq:EA-pm-quantico}
           p(b \vert x, y) = \Tr[(\mathcal{C}_x \otimes \mathcal{I})(\rho_{AB}) \cdot M_{b \vert y}] 
        \end{equation}
        for all $x \in [n_{X}], y \in [n_{Y}]$ and $b \in [n_{B}]$. 
    \end{defi}
    For more specific instances, we denote the following set of behaviors: The case when the shared resource $\rho_{AB}$ is a specific state $\rho$, we denote $Q_{d}^{\rho}$, and by $Q_{d}^{E}$ we mean that $\rho_{AB}$ has any finite dimension. Clearly, then,
    \begin{equation}
        Q_{d}^{D} \subseteq Q_{d}^{E}.
    \end{equation}

    As for the classical communication case, one can show the definition to be equivalent to a three-step procedure where (i) Alice performs an $d$-outcome measurement on her share of $\rho_{AB}$, then (ii) she communicates the result of her measurement to Bob over a classical channel, and finally (iii) based on his input $y$ and the message received, Bob performs a measurement on his part of $\rho_{AB}$ \cite{tavakoli2021correlations,pauwels2021entanglement}. More precisely (see also Fig. \ref{fig:ea-cpm}),
    \begin{defi}[Entanglement-assisted classical behaviors]\label{def:EA-pm-classico}
        A behavior $\{ p(b \vert x, y) \}_{b, x, y} $ belongs to $C_{d}^{D}(n_{X}, n_{Y}, n_{B})$ if there exists (i) a state $\rho_{AB} \in \mathcal{L}(\mathbb{C}^{D} \otimes \mathbb{C}^D)$, (ii) for each x in $[n_{X}]$, a set of d-outcome POVMs $\{M_{a|x} \}_{a = 0}^{d-1}$ on $\mathcal{L}(\mathbb{C}^{D})$, and (iii) for each $y \in [n_{Y}]$ and $a \in [d]$, a set of $n_{B}$-outcome POVMs $\{ N_{b \vert a, y} \}_{b \in [n_{B}]}$, such that
            $$ p(b \vert x, y) = \sum_{a=0}^{d-1}\Tr[\rho_{AB} (M_{a|x} \otimes N_{b \vert a, y})],$$
        for all $x \in [n_{X}], y \in [n_{Y}]$ and $b \in [n_{B}]$.
    \end{defi}

    Correspondingly, we specify $C_{d}^{\rho}$ and $C_{d}^{E}$ as before, wherefrom
    \begin{equation}
       C_{d}^{D} \subseteq C_{d}^{E}.
    \end{equation}

    As mentioned, several important questions remain open in entanglement-assisted prepare-and-measure scenarios. Some examples are determining the cost of classically simulating a quantum behavior, showing whether higher-dimensional entanglement is always associated with greater advantages and understanding if every type of entangled resource is useful. These are the questions we answer in the following section.

\section{Results}
\label{sec:results}

    \subsection{Classical versus quantum communication in entanglement-assisted prepare-and-measure scenarios}
    \label{sec:c-vs-q-comm}
    
        A central objective in \textsc{ea-pm} scenarios is to understand the relationships between quantum and classical communication or, more precisely, between the sets $C_{d}^{D}$ and $Q_{d'}^{D'}$. In particular, even when the channel capacity of the two sets is the same, non-trivial relationships can be found.
        
        The cases involving bits and qubits are by far the most explored \cite{tavakoli2021correlations, pauwels2021entanglement, Frenkel2022entanglement, Pauwels_adaptive_2022}.
        However, few results for systems of arbitrary dimension are found in the literature. Our next result goes directly in that direction, showing a non-trivial family (in the sense that both sets have the same channel capacity) of inclusions between the \textsc{ea} classical behaviors with 2 dits of communication and \textsc{ea} quantum behaviors with a single qudit of communication.
        \begin{restatable}[Behavior simulation cost]{res}{restwo}\label{result: C_d^2 = Q_d}
            For any choice of $D, d, n_{X}, n_{Y}, n_{B}$ we have 
            \begin{equation}
              C_{d^2}^{D}(n_{X}, n_{Y}, n_{B}) \subseteq Q_{d}^{D \cdot d}(n_{X}, n_{Y}, n_{B}),  
            \end{equation}
            and 
            \begin{equation}
                Q_{d}^{D}(n_{X}, n_{Y}, n_{B}) \subseteq C_{d^2}^{D \cdot d}(n_{X}, n_{Y}, n_{B}).
            \end{equation}
        \end{restatable}
        \noindent This result shows that in \textsc{ea-pm} scenarios, with an appropriate amount of additional entanglement, we can always send a single qubit to simulate the behaviors obtained by sending 2 classical dits, and \textit{vice versa}. The proof of this result, shown in appendix \ref{sec: appendix proof C_d^2 = Q_d}, relies on the duality between quantum teleportation \cite{bennett1993teleporting} and dense-coding protocols \cite{bennet1992communication, Werner_2001_All}.
        
        A direct application of Result \ref{result: C_d^2 = Q_d} leads us to the main result of this section.
        \begin{res}[Equivalence of classical/quantum behaviors]\label{result:C_d^2=Q_d}
            For any choice of $d, n_{X}, n_{Y}, n_{B}$ we have 
            \begin{equation}
                 C_{d^2}^{E}(n_{X}, n_{Y}, n_{B}) = Q_{d}^{E}(n_{X}, n_{Y}, n_{B}).
            \end{equation}
            
        \end{res}
        \noindent Thus, if entanglement is a free resource between the parties, in the paradigm of \textsc{pm} scenarios, the communication of one quantum dit is equivalent to the communication of two classical dits. This may seem like a trivial consequence of dense-coding \cite{bennet1992communication}, however, it is worth emphasizing that the result presented here concerns the whole sets of correlations, and not only the amount of information transmitted. From the perspective of Eq.~\eqref{eq:result-Frenkel}, it is also remarkable that, here, equality holds for any $(n_X, n_Y, n_B)$ setting, while in Ref.\@ \cite{frenkel-storage-2015} 
        sets of behaviors were equal only for restricted settings. Perhaps even more surprisingly, as shown in \cite{tavakoli2021correlations} and generalized in the next section, dense coding is not the optimal protocol for certain communication tasks in \textsc{ea-pm} scenarios.

    \subsection{Advantage in higher-dimensional entanglement assistance for every dimension}
    \label{sec:highdim}
    
        In a seminal paper, Bennet \textit{et al}.~\cite{bennet1992communication} showed that if Alice and Bob share a maximally entangled pair of qudits, the communication of a single qudit from Alice to Bob can transmit two dits of information.
        This so-called dense coding protocol may seem to contradict Holevo's bound \cite{holevo1973bounds}, since by sending a single qubit one can transmit two bits of information. But, on a closer inspection, it only reveals that Holevo's bound does not apply to situations where the parties share quantum correlations.
        
        An immediate question is whether with an arbitrary amount of entanglement it is possible for Alice to transmit more information than two dits. Unfortunately, as shown in \cite{nayak2002communication}, this is not the case. In this context, for the task of guessing Alice's input, the dense coding protocol is optimal and the capacity of a quantum channel can be at most doubled under the presence of entanglement.  
        
        Despite the fact that dense coding is optimal at the quantum channel capacity level, it is not necessarily the optimal strategy at the correlation level. As it has recently been shown, $Q_2^2$ is a strict subset of $Q_2^4$ \cite{tavakoli2021correlations}. In other words, for scenarios where communication is limited, a maximally entangled state with the same local dimension as the communicated quantum system ($Q_2^2$) does not provide the whole set of behaviors $Q_2^E$.
        Therefore, even though high-dimensional entanglement does not increase a qudit's information capacity, it can be a useful resource in other communication tasks, where  by ``high dimensional entanglement'' we mean states whose local dimension is higher than that of the communicated quantum system. 
        
        In this section, we will extend the $Q_{2}^{2} \subsetneq Q_{2}^{4}$ result, shown in Ref.~\cite{tavakoli2021correlations}, to arbitrary dimensions, solving one of the open problems listed in Ref.~\cite{pauwels2021entanglement}. More precisely, we are going to show that higher-dimensional entanglement is useful irrespective of the dimension of the message. 
        
        Before getting to that, let us show that an entangled pair of qubits is useful for classical communication, regardless of the size of the message. To do so, we first consider the $C_{2}(3,1,4)$ scenario. By listing all of its extremal points and transforming to a hyperplane representation \cite{lorwald2015panda}, we obtain the inequalities defining the polytope of $\bar{C}_{2}(3,1,4)$ behaviors. The only nontrivial inequality thus obtained is
        \begin{equation}\label{eq: Desigualdade F_2}
           \mathcal{F}_{2}[\textbf{p}(b|x)] =  2\sum_{i=0}^{2} p(i|i) + \sum_{i=0}^{2}p(3|i) \le 4.
        \end{equation}
        The linear function $\mathcal{F}_2$ can be interpreted as a game where Alice is given an input $x$ out of three possible inputs, $x \in [3]$. Alice sends a bit to Bob, and Bob must provide an answer $b \in [4]$. If Bob's answer is the same as Alice's input, i.e., $b=x$, then they get two points in the game. However, Bob also has an extra alternative answer, the output $b=3$, which gives them
        one point regardless of what Alice's input was.
        
        We note that the game is not trivial, since Alice is limited to sending only one bit to Bob, thus Bob cannot know what Alice's input was in every round. We can view the distribution $p(b|x)$ as the probability that Bob gives $b$ as an answer, when Alice receives $x$ as input. Therefore, the linear function \eqref{eq: Desigualdade F_2} can be viewed as the average number of points that Alice and Bob receive when they play the game following the strategy $\textbf{p}(b|x)$ (in Ref.~\cite{Doolittle_2021_Certifying}, similar \textit{ambiguous guessing games} were introduced).
        
        Inequality \eqref{eq: Desigualdade F_2} is a facet of the polytope $\bar{C}_{2}(3,1,4)$, so no classical behavior [the ones in $\bar{C}_2(3,1,4)$] can violate it \cite{repo}. Nevertheless, the introduction of entanglement-assistance enables violations. In fact, by employing the alternated optimization procedure developed in \cite{tavakoli2021correlations}, one can find a behavior  $\mathbf{p^\prime}(b|x) \in C_{2}^{2}(3,1,4)$ such that (see \cite{repo}) 
        \begin{equation}\label{eq: max C_4^2}
            \mathcal{F}_{2}[\mathbf{p^\prime}(b|x)] = 4.155.
        \end{equation}

        To generalize this result, let us extend this inequality to scenarios $C_{d}(d+1,1,d+2)$ by defining the linear functional
        \begin{equation}\label{eq: Desigualdade F_d}
            \mathcal{F}_{d}[\textbf{p}(b|x)] =  2\sum_{i=0}^{d} p(i|i) + \sum_{i=0}^{d}p(d+1|i),
        \end{equation}
        which can be interpreted analogously to $\mathcal{F}_2$. In Appendix \ref{sec:Bounds-Fd}, we explicitly describe a $C_d$ strategy that can reach $\mathcal{F}_{d} = 2d$. In fact, as also shown therein, $2d$ is the maximum value of $\mathcal{F}_{d}$ for all behaviors in $\bar{C}_{d}$. On top of that, we also show that it is always possible to obtain a value for $\mathcal{F}_d$ greater than $2d$ when considering the entanglement-assisted behaviors in $C_{d}^{2}$. This implies that there is always a $C_d^2$ behavior which is not in $C_d$ or, in other words, that
        \begin{res}\label{res: C_d < C_d^2}
            For every integer $d\ge 2$, $C_{d} \subsetneq C_{d}^{2}$.
        \end{res}
        
        A simple concatenation of previous results proves the main claim of this section, i.e., that higher-dimensional entanglement is a useful resource regardless of the communication dimension.
        \begin{res}
            For every $d \ge 2$, $\bar{Q}_{d}^{d} \subsetneq \bar{Q}_{d}^{d+2}.$
        \end{res}
        \begin{proof}
            We first notice that $\bar{C}_{d^2} \subseteq \bar{Q}_{d}^{d}$ (this inclusion follows directly from the dense coding protocol). On the other hand, it is also straightforward that $\bar{Q}_{d}^{d} \subseteq \bar{Q}_{d^2}$. So, $\bar{C}_{d^2} \subseteq \bar{Q}_{d}^{d} \subseteq \bar{Q}_{d^2}$. Focusing on scenarios with $n_{Y} = 1$, where, by \eqref{eq:result-Frenkel}, $\bar{C}_{d^2} = \bar{Q}_{d^2}$, we get
            \begin{equation}\label{eq: Q_d^d = C_{d^2}}
                \bar{Q}_{d}^{d} = \bar{C}_{d^2}.
            \end{equation}
            Then, combining Eq.~\eqref{eq: Q_d^d = C_{d^2}} with Result \ref{res: C_d < C_d^2} and Result \ref{result: C_d^2 = Q_d},
            \begin{equation*}
                \bar{Q}_{d}^{d} = \bar{C}_{d^2} \subsetneq \bar{C}_{d^2}^{2} \subseteq \bar{Q}_{d}^{d+2}.
            \end{equation*}
        \end{proof}
    
    \subsection{Not all entangled states are useful resources for classical communication tasks}
    \label{sec:steering}
    
        A related, also pivotal question in \textsc{ea-pm} scenarios is to find which types of shared correlations are useful resources. This question was raised, for instance, in \cite{pauwels2021entanglement}, in which the authors ask whether for any entangled state $\rho$ there exists a $d$ such that $\bar{C}_{d}^{\rho} \not\subseteq \bar{C}_{d}$. Naturally, the analogue for quantum communication scenarios, i.e., whether $\bar{Q}_{d}^{\rho} \not\subseteq \bar{Q}_{d}$, is also of interest.
        
        When $\rho$ is a pure state, both results are true \cite{Cleve_Substituting_1997, Trojek_Experimental_2005, Tavakoli2020doesviolationofbell, Patra_classical_dense_coding_2022}. For instance, in \cite{Patra_classical_dense_coding_2022}, the authors present a game (4-cup $\&$ 2-ball game) in the scenario $C_{2}(6,1,4)$ where the advantage brought by an entanglement-assistance $\rho$ grows at least linearly w.r.t.\@ $\rho$'s violation of a \textsc{chsh} inequality \cite{CHSH-inequality}. As every entangled pure state violates some \textsc{chsh} inequality \cite{Gisin_Bell's_1991, Yu_all_entangled}, then every entangled pure state provides an advantage in this game.
        
        Similarly, for quantum communication, it has been shown that the optimal probability of success of Bob guessing two of Alice's bits, with Alice sending a single qubit but assisted by an entangled state, increases with the Schmidt number of the shared state between them \cite{Moreno_Semi_device_2021}.
        
        This is no longer the case when we consider mixed states in classical communication \textsc{ea-pm} scenarios, since it is then possible to find entangled states which are not useful resources. To prove so, we will show that all \textit{unsteerable} states lead to classical behaviors, therefrom uncovering a tight connection between \textit{steering} and prepare-and-measure scenarios.
        
        Let us start introducing the ideas behind quantum steering \cite{schrodinger_discussion_1935, Wiseman_Steering_2007,Uola_Quantum_Steering_2020}. An \textit{assemblage} $\{\varrho_{a|x}\}$ is a collection of ensembles for a same state $\rho_{B}$, i.e., $\sum_{a}\varrho_{a|x} = \rho_B, \forall x$. For a fixed bipartite state $\rho_{AB}$, every collection of measurements $\{M_{a|x}\}$ performed locally by Alice leads to an assemblage on Bob's side described by the elements
        \begin{equation}\label{eq:Assemblage-from-state}
            \varrho_{a|x} = \Tr_{A}[\rho_{AB}(M_{a|x} \otimes \mathbb{1})].
        \end{equation}
        
        An assemblage $\{\varrho_{a|x}\}$ is \emph{unsteerable} (from Alice to Bob) \footnote{In this manuscript, the concept of steering will always be for the case of Alice for Bob. This asymmetry comes from the fact that in a prepare-and-measure scenario, Alice and Bob play a different role, Alice performs the preparation and Bob the measurement.}, if it admits a local hidden state model, that is, if each element can be written as
        \begin{equation}\label{eq:LHS-model}
            \varrho_{a|x} = \sum_{\lambda}p(\lambda)p(a|x,\lambda)\sigma_{\lambda},
        \end{equation}
        with $\sigma_\lambda \in \mathcal{L}(\mathcal{H}_B)$ and $\{ p(\lambda) \}$, $\{ p(a|x,\lambda) \}_{x,\lambda}$ being probability distributions \cite{Wiseman_Steering_2007}.
        
        More fundamentally, a state $\rho_{AB}$ is unsteerable if for \emph{every} possible choice of measurements $\{M_{a|x}\}$, the assemblage given by Eq.~\eqref{eq:Assemblage-from-state} is unsteerable. Otherwise, $\rho_{AB}$ is said to be steerable. Interestingly, there are entangled states, such as some Werner and isotropic states, which are not steerable for any set of \textsc{povm}s \cite{quintino-inequivalence-2015,chau_steering_povms}.
        
        So let us suppose that $\rho_{AB}$ is an unsteerable state. As stated in Definition \ref{def:EA-pm-classico}, a behavior belongs to $C_{d}^{\rho}$ if
        \begin{equation}
            p(b \vert x, y) = \sum_{a=0}^{d-1}\Tr[\rho_{AB} (M_{a|x} \otimes N_{b \vert a, y})].
        \end{equation}
        From a well-known property of the trace, $\Tr_{A}[X_{AB}(\mathbb{1}_A \otimes Y_B)] = \Tr_{A}(X_{AB})Y_B$, we get that
        \begin{align}
            p(b \vert x, y) &= \sum_{a=0}^{d-1}\Tr\left\{ \Tr_{A}[\rho_{AB} (M_{a|x} \otimes \mathbb{1}_B)] N_{b \vert a, y}\right\} \nonumber\\
            &= \sum_{a=0}^{d-1}\Tr\left( \varrho_{a|x} N_{b \vert a, y}\right).
        \end{align}
        With any unsteerable $\rho_{AB}$ [Eq.~\eqref{eq:LHS-model}],
        \begin{align}\label{eq:22}
            p(b \vert x, y)
            &= \sum_{a=0}^{d-1}\sum_{\lambda}p(\lambda)p(a|x,\lambda)\Tr\left( \sigma_{\lambda} N_{b \vert a, y}\right)\nonumber\\
            &= \sum_{a=0}^{d-1}\sum_{\lambda}p(\lambda)p(a|x,\lambda)p(b|a,y,\lambda) ,
        \end{align}
        which is the definition of a behavior belonging to $\bar{C}_{d}$ (see Eq.~\eqref{eq:behaviour_classical_SR}). Thus, any unsteerable resource $\rho$ leads to $\bar{C}_{d}^{\rho}(n_{X}, n_{Y},n_{B}) = \bar{C}_{d}(n_{X}, n_{Y},n_{B})$, for all choices of $d, n_X, n_Y$ and $n_B$. As not all entangled states are steerable \cite{chau_steering_povms}, we conclude that
        \begin{res}\label{res:steering}
            Some entangled states do not provide advantages in classical communication scenarios.
        \end{res}
        Interestingly, while steering is necessary for classical communication advantages, it is not when communication is quantum (see e.g.\@ Result 3 in \cite{Moreno_Semi_device_2021}).
        
        Having connected steering with prepare-and-measure scenarios, some further results from the former readily translate to the latter. For instance, it is well known that, for an assemblage to be steerable, Alice must have used incompatible (in the sense of non-jointly measurable) measurements \cite{MT_Joint_2014, Uola_Joint_2014}. In light of Definition \ref{def:EA-pm-classico}, it immediately follows that to have advantages in entanglement-assisted classical communication tasks, Alice's measurements have to be incompatible. Naturally, these observations can be applied as semi-device independent witnesses for steering and measurement incompatibility.
        
        It is important to call attention to the results of Ref.~\cite{Jebarathinam_Superunsteerability_2019}. In there, it is shown that unsteerable states can be useful for \textsc{ea-rac} in the situation where the amount of randomness shared between the parties is bounded. Note that this does not contradict Result \ref{res:steering}, as we are not bounding the amount of shared randomness.

\section{Discussion}

In this manuscript, we investigated aspects of quantum and classical communication tasks under the prepare-and-measure (\textsc{pm}) scenario paradigm.
First, we formally showed that measurement incompatibility is necessary for quantum communication to outperform classical communication. It is widely accepted that this result is a direct corollary of a result from \cite{frenkel-storage-2015}, however, to the best of our knowledge, no proof of it was presented so far.

We then focused on \textsc{pm} scenarios assisted by entangled quantum states --- a setting that has only recently begun to be systematically investigated \cite{tavakoli2021correlations,pauwels2021entanglement}. In this context, we derived a chain of inclusions between classical and quantum sets of behaviors by increasing only the dimension of the assisted entanglement. This led to the observation that, when arbitrary entanglement is available, the sets of quantum and classical communication behaviors are identical. We furthermore showed that there are classical communication protocols in which certain entangled pairs of qubits always lead to better performance, no matter what is the communication dimension.

Subsequently, by employing our chain of inclusions between classical and quantum sets of behaviors, we proved that increasing the entanglement dimension always leads to better performance, no matter the communication dimension, solving an open problem stated in Ref.~\cite{pauwels2021entanglement}.

Lastly, we discussed which properties entangled states must have to provide advantages in classical communication scenarios, showing that unsteerable states are useless in such tasks. Connecting this with previous results in quantum steering \cite{chau_steering_povms}, we concluded that not all entangled states are useful resources in classical communication, this being another of the open problems listed in Ref.~\cite{pauwels2021entanglement}.

Several connections between ours and previous results can be highlighted. Firstly, Pauwels \textit{et al.} \cite{pauwels2021entanglement} questioned whether there is an integer $D$ such that $C_{d}^{D} = C_{d}^{D+k}$ for all $k$, and analogously for quantum communication, i.e., $Q_{d}^{D} = Q_{d}^{D+k}$ for all $k$. From Result \ref{result: C_d^2 = Q_d}, we see that these two questions are equivalent. Indeed, answering one, our result will guarantee the same answer to the other. This can be of great help, as the case with classical communication can be more easily linked to Bell scenarios, in which we know that an arbitrary amount of entanglement is useful \cite{Vertesi_maximal_2010, slofstra_set_2019, coladangelo_inherently_2020}.

Secondly, in Ref.~\cite{tavakoli2021correlations}, two algorithms are developed: one to bound the correlations in $C_{d}^{E}$, and another for the ones in $Q_{d}^{E}$. However, from Result \ref{result:C_d^2=Q_d}, $C_{d^2}^{E} = Q_{d}^{E}$, therefore we conclude that both algorithms solve the same problem. 

Lastly, Frenkel and Weiner \cite{Frenkel2022entanglement} showed that, for communication channel scenarios (i.e., \textsc{pm} scenarios with $n_Y = 1$), $C_{2}^{D^*} \subseteq C_{4}$ for every $D$ (The superscript $D^*$ represents the assistance of a maximally entangled state of local dimension $D$.). They also conjecture that $C_{d}^{E} \subseteq C_{d^2}$ for every $d$. If the conjecture ends up being true, we can readily apply Results \ref{result:C_d^2=Q_d}, along with \eqref{eq:result-Frenkel}, to conclude that $Q_{d}^{E} \subseteq Q_{d^4}$ for every $d$. This can also be seen as an alternative way to prove this conjecture.

Further questions naturally arise from our results. For one, it would be interesting to find a characterization for the set of entangled states that are useful in \textsc{ea-pm} scenarios with classical communication. As discussed, this is a subset of the steerable states. The question remains whether it is a strict subset or not, and if so, what is the relationship between this set and the set of nonlocal states? It would also be interesting to find closer connections between measurement incompatibility and prepare-and-measure scenarios. 

\begin{acknowledgments}
    Many thanks to Marcelo Terra Cunha, Armin Tavakoli and Jef Pauwels for all the discussions and suggestions. C.G.\@ acknowledges support from FAEPEX/UNICAMP (Grant No. 3044/19), the Deutsche Forschungsgemeinschaft (DFG, German Research Foundation, project numbers 447948357 and 440958198), the Sino-German Center for Research Promotion (Project M-0294), and the House of Young Talents of the University of Siegen. This work was also supported by the Brazilian National Council for Scientific and Technological Development (CNPq), the National Institute for Science and Technology on Quantum Information (INCT-IQ) (Grant No. 465469/2014-0), and the São Paulo Research Foundation FAPESP (Grant No. 2018/07258-7).
\end{acknowledgments}
\vspace{0.5cm}

\noindent\textit{Note added.—}In Ref.~\cite{Pauwels_adaptive_2022}, a stronger version of Result \ref{res:steering} is presented, namely, that no local state can provide advantages in classical communication tasks. It is worth mentioning, however, that the preprint version of this manuscript predates Ref.~\cite{Pauwels_adaptive_2022}, and has been properly cited in the context of the aforementioned stronger result.

\bibliography{references}

\newpage
\onecolumngrid
\appendix

\section{Measurement incompatibility is necessary for nonclassicality}\label{sec:appendix-proof-incompatibility-pm}

    In general, for prepare-and-measure scenarios without entanglement assistance, $Q_{d}(n_{X}, n_{Y}, n_{B}) \nsubseteq C_{d}(n_{X}, n_{Y}, n_{B})$. Here we show that, when Bob's measurements are jointly measurable, these sets of behaviors become equal.

	Let $p(b|x,y) \in Q_{d}(n_{X}, n_{Y}, n_{B})$ be given by $p(b|x,y) = \Tr(\rho_x M_{b|y})$, where the collection of Bob's POVMs we suppose to be jointly measurable. Given that, there exists a \textit{mother} POVM 
    $$\mathcal{N} = \{ N_{z_1,\ldots,z_{m}} \mid z_{1},\ldots,z_{m} \in [n_B]\},$$
    such that %
        \begin{equation}
            M_{b\vert y} = \sum_{\substack{z_1,\ldots,z_{m}\\z_y = b}} N_{z_1,\ldots,z_{m}}, \ \ \forall \ b,y.
        \end{equation}
	where for simplicity, we are denoting $m = n_Y$. Hence, the observed statistics are given by
	\begin{equation}\label{eq:behavior_grand}
		p(b \vert x,y) = \Tr(\rho_{x}M_{b\vert y}) =  \sum_{\substack{z_1,\ldots,z_{m} = 0 \\z_y = b}}^{n_B - 1} \Tr(\rho_{x} N_{z_1,\ldots,z_{m}}).
	\end{equation}
	Let's define $\bm{q}$ as the behavior:
	\begin{equation}
		q(z_1,\cdots,z_{m} | x) = \Tr(\rho_{x} N_{z_1,\ldots,z_{m}}) \ \ \forall z_{1}, \cdots, z_{m} \ \in \ [n_B].
	\end{equation}
	It follows that $\bm{q}$ is a behavior in $Q_{d}(n_{X}, 1, k)$ where $k =  (n_{B})^{n}$. 
    We may regard the result from \cite{frenkel-storage-2015}, priorly mentioned in Eq.~\eqref{eq:result-Frenkel}, which states that $\bar{C}_{d}(n_X,1,k) = \bar{Q}_{d}(n_X,1,k)$, that is, in a scenario where Bob has a single measurement, there is always a classical model that simulates the statistics attainable with quantum resources. Thus, by the definition
	\begin{align}\label{eq:Born_grand_POVM}
		\Tr(\rho_{x} N_{z_1,\ldots,z_{m}}) &= q(z_1,\ldots,z_{m} \vert x) \nonumber\\
		&= \sum_{a=0}^{d-1}\sum_{\lambda} \pi(\lambda) q_{A}(a\vert x,\lambda)q_{B}(z_1,\ldots,z_{m}\vert a,\lambda),
	\end{align}
	for all $x \in [n_X]$ and $z_1,\ldots,z_{m} \in [n_B]$. 
	Substituting \eqref{eq:Born_grand_POVM} in \eqref{eq:behavior_grand}, we have:
	\begin{align}\label{eq:behavior_grand2}
		p(b \vert x,y) &= \sum_{\substack{z_1,\ldots,z_{m} = 0 \\z_y = b}}^{n_B - 1} \Tr(\rho_{x} N_{z_1,\ldots,z_{m}}) \nonumber\\
		&= \sum_{\substack{z_1,\ldots,z_{m} = 0 \\z_y = b}}^{n_B - 1} \sum_{a=0}^{d-1} \sum_{\lambda} \pi(\lambda) q_{A}(a \vert x,\lambda) q_{B}(z_1,\ldots,z_{m} \vert a,\lambda).
	\end{align}
	Now, define
	\begin{equation}\label{eq:Classical_model_Bob}
		q_{B}(b \vert a, y, \lambda) =  \sum_{\substack{z_1,\ldots,z_{m} = 0 \\z_y = b}}^{n_B - 1} q_{B}(z_1,\ldots,z_{m} \vert a,\lambda),
	\end{equation}
	for all $b,a,y,\lambda$. It is easy to see that $q_{B}(b \vert a,y,\lambda)$ are probability distributions. Substituting  \eqref{eq:Classical_model_Bob} in \eqref{eq:behavior_grand2}, we have:
	%
	\begin{align*}
		p(b \vert x,y) &= \sum_{\substack{z_1,\ldots,z_{m} = 0 \\z_y = b}}^{n_B - 1} \sum_{a=0}^{d-1} \sum_{\lambda} \pi(\lambda) q_{A}(a \vert x,\lambda) q_{B}(z_1,\ldots,z_{m} \vert a,\lambda)\\
        &= \sum_{a=0}^{d-1} \sum_{\lambda} \pi(\lambda) q_{A}(a \vert x,\lambda)\left[ \sum_{\substack{z_1,\ldots,z_{m} = 0 \\z_y = b}}^{n_B - 1} q_{B}((z_1,\ldots,z_{m}) \vert a,\lambda) \right] \\
        &=  \sum_{a=0}^{d-1} \sum_{\lambda} \pi(\lambda) q_{A}(a \vert x,\lambda) q_{B}(b \vert a,y,\lambda).
	\end{align*}
	%
	Therefore, $p(b\vert x,y)$ can be reproduced by preparing, sending, and then measuring classical instead of quantum states. Hence we conclude that measurement incompatibility is necessary for observing nonclassicality in \textsc{pm} scenarios.

\section{Higher-dimensional Bell basis}
    First, we review the generalization of Bell basis and Pauli matrices to higher-dimensional Hilbert spaces, which will become handy for proving Result \ref{result: C_d^2 = Q_d}.
    
    In order to generalize the quantum teleportation protocol to higher-dimensional Hilbert spaces $\mathcal{H}_A^\prime \otimes \mathcal{H}_B^\prime$ (let us suppose that $\mathcal{H}_A'$ and $\mathcal{H}_B'$ has the same dimension), Bennett \textit{et al.}~\cite{bennett1993teleporting} introduced an orthogonal basis of maximally entangled states  $\ket{\psi_{nm}}$, $0\le n,m \le d-1$. For this, let $\{\ket{i}_A'\}$ and $\{\ket{i}_B'\}$ be orthonormal bases of $\mathcal{H}_A'$ and $\mathcal{H}_B'$, respectively, and define the states
    \begin{align}
    	\ket{\phi_{nm}} = \frac{1}{\sqrt{d}}\sum_{j=0}^{d-1} e^{2\pi i j n / d} \ket{j}_{A'}\otimes \ket{j \oplus_{d} m}_{B'}, \label{Eq:BellStates}
    \end{align}
    where $j \oplus_d m = (j+m)\,\text{mod }d$. The states $\ket{\phi_{nm}}$ are easily seen to be orthonormal,
    \begin{align*}
    	\braket{\phi_{n' m'}}{\phi_{n m}} &=
    	\frac{1}{d} \sum_{j,k=0}^{d-1} e^{2\pi i (j n - k n') / d} \braket{k}{j}_{A'} \braket{k \oplus_{d} m'}{j \oplus_{d} m}_{B'} \\
    	&= \frac{1}{d} \sum_{k=0}^{d-1} e^{2\pi i k (n - n') / d} \  \delta_{m,m'} = \delta_{n,n'} \delta_{m,m'} .
    \end{align*}

    To generalize the Pauli matrices, Bennett et al.~\cite{bennett1993teleporting} introduced the unitary operators 
    \begin{align}\label{Eq:Unm}
    	U_{nm} := \sum_{k=0}^{d-1} e^{2 \pi i k n / d} \ket{k}_{A'}\bra{k \oplus_{d} m}_{A'}.
    \end{align}
    Wherefrom we can see that
    \begin{equation}\label{Eq:Unm-phi}
        U_{nm} \ket{\phi_{00}} = \ket{\phi_{nm}},
    \end{equation}
    for all $0\le n,m \le d-1$.

\section{Proof of Result \ref{result: C_d^2 = Q_d}}\label{sec: appendix proof C_d^2 = Q_d}

    \restwo*
    \begin{proof}
        Let us start with the inclusion $C_{d^2}^{D} \subseteq Q_{d}^{D \cdot d}$. Given a behavior in $C_{d^2}^{D}$ with elements $p(b \vert x,y)$, by Definition \ref{def:EA-pm-classico}, there exist (i) a quantum state $\rho_{AB} \in \mathcal{L}(\mathbb{C}^{D} \otimes \mathbb{C}^{D})$; (ii) for each $x \in [n_{X}]$, a set of POVMs $\mathcal{M}_{x} = \{M_{k,l|x} \}_{k,l \in [d]}$ \footnote{we are using two dits to represent the $d^2$ possible results.}; and (iii),  for each $k,l \in [d]$ and $y \in [n_{Y}]$ a POVM $\{N_{b|k,l,y}\}_{b \in [n_{B}]}$ such that
        \begin{equation}\label{Eq:p(b|x,y) in C}
            p(b|x,y) = \sum_{k,l = 0}^{d-1} \Tr\left[ \rho_{AB}(M_{k,l|x} \otimes N_{b|k,l,y})\right].
        \end{equation}
        We are looking for a realization of this behavior in $Q_{d}^{D \cdot d}$. To find it, let us suppose that Alice and Bob share the following entangled state $\rho_{AB} \otimes \ketbra{\phi_{00}}{\phi_{00}}_{A'B'}$, where $\ket{\phi_{00}}_{A'B'} = \frac{1}{\sqrt{d}}\sum_{i=1}^{d^2} \ket{ii}$ is a maximally entangled state (see Eq.~ \eqref{Eq:BellStates}). Let us assume further that systems $A$ and $A'$ are in Alice's lab, while systems $B$ and $B'$ are in Bob's lab. To simulate the behavior $p(b|x,y)$, they proceed with the following strategy.
        \begin{enumerate}
            \item For each $x \in [n_{X}]$, Alice performs measurement $\{M_{k,l|x}\}_{k,l=0}^{d-1}$ on her subsystem $A$.
            She gets the pair of dits $k,l$ as a result, with probability $\Tr(\rho_{AB} (M_{k,l|x} \otimes \mathbb{1}))$.
           
            \item Alice wants to send her measurement result to Bob. For that, Alice and Bob will perform the dense coding protocol, using the state $\ket{\phi_{00}}$ that they both share. Suppose then that the result of Alice's measurement was the pair of dits $r,s$. Alice  then applies the unitary $U_{r,s}$ (Eq.~\eqref{Eq:Unm}) over her share of the state $\ket{\phi_{00}}$, resulting in the state $\ket{\phi_{rs}}$ (Eq.~\eqref{Eq:Unm-phi}). Since she is allowed to communicate a $d$-dimensional quantum system, she sends her qudit to Bob.
            \item Bob applies a projective measurement on the basis $\{\ket{\phi_{kl}}\}_{k,l=0}^{d-1}$ on the state $\ket{\phi_{rs}}$, thus obtaining the dit pair $r,s$ as the result.
            \item Finally, he performs the POVM measurement $\{N_{b|r,s,y}\}_{b \in [n_{B}]}$ on his part of the state $\rho_{AB}$, where $y$ is his choice of input, obtaining outcome $b$.
        \end{enumerate}
        Using Bayes' rule, we can write the behavior originating from this procedure as:
        \begin{align*}
            p(b|x,y) &= \sum_{r,s = 0}^{d-1} p(r,s|x)p(b|r,s,y) \\
            &=  \sum_{r,s = 0}^{d-1}\Tr\left[ \rho_{AB}(M_{r,s|x} \otimes \mathbb{1})\right] \cdot
            \Tr\left\{ \frac{\Tr_A\left[ \rho_{AB}(M_{r,s|x} \otimes \mathbb{1})\right]}{\Tr\left[ \rho_{AB}(M_{r,s|x} \otimes \mathbb{1})\right]} N_{b|r,s,x} \right\} \\
            &= \sum_{r,s = 0}^{d-1}\Tr\left\{ \Tr_A[ \rho_{AB}(M_{r,s|x} \otimes \mathbb{1})] N_{b|r,s,x}\right\} \\
            &= \sum_{r,s = 0}^{d-1}\Tr\left[ \rho_{AB}(M_{r,s|x} \otimes N_{b|r,s,x})\right],
        \end{align*}
        which recovers exactly the same behavior as Eq.~\eqref{Eq:p(b|x,y) in C}.
        
        We are left with the inclusion $Q_{d}^{D} \subseteq C_{d^2}^{D \cdot d}$. From Definition \ref{def: EA-pm-quantico}, given $\{ p(b \vert x, y) \} \in Q_{d}^{D}$ there exist (i) a state $\rho_{AB} \in \mathcal{L}(\mathbb{C}^{D} \otimes \mathbb{C}^{D})$; (ii) a set of CPTP maps $\{\mathcal{C}_x \}_{x\in [n_{X}]}$, with $\mathcal{C}_x: \mathcal{L}(\mathbb{C}^{D}) \to \mathcal{L}(\mathbb{C}^{d})$; and (iii) a set of quantum measurements $\mathcal{M}_y$, where each $\mathcal{M}_y = \{ M_{b \vert y} \}_{b}$ is a POVM, such that
        $$
            p(b \vert x, y) = \Tr\left[ (\mathcal{C}_x \otimes \mathcal{I})(\rho_{AB}) M_{b \vert y} \right].
        $$
        
        To show that such behavior also belongs to $C_{d^2}^{D \cdot d}$, it is sufficient to show that Alice and Bob can remotely prepare the states $\varrho_{AB}^{x} = (\mathcal{C}_x \otimes \mathcal{I})(\rho_{AB})$ in Bob's lab, using only two dits of communication assisted by the entangled state $ \rho_{AB} \otimes \ketbra{\phi_{00}}{\phi_{00}}_{A'B'} \in \mathcal{L}(\mathbb{C}^{D \cdot d} \otimes \mathbb{C}^{D \cdot d})$.
        
        Let us start by fixing orthogonal bases for the spaces $A$ and $B$, given by $\{\ket{j}_A\}_{j=0}^{d-1}$ and $\{\ket{k }_B\}_{k=0}^{D-1}$, respectively. For simplicity, let us initially assume that $\varrho_{AB}^{x} $ is pure, i.e., $\varrho_{AB}^{x} = \ketbra{\varphi_x}{\varphi_x}_{AB}$. Since $\{\ket{j}_A \otimes \ket{k}_B\}$ is a basis of the space $A \otimes B$, there is a set of coefficients $\{c_{j,k}\} \subset \mathbb{C}$ such that
        $$
            \ket{\varphi_x}_{AB} = \sum_{j=0}^{d-1}\sum_{k=0}^{D-1} c_{jk}\ket{j}_A\ket{k}_B.
        $$
        Thus, the shared state between Alice and Bob after the application of the channel $\mathcal{C}_{x}$ on the systems $A$ and $B$ is given by
        $$
            \ket{\varphi_x}_{AB} \ket{\phi_{00}}_{A'B'} = \sum_{j,l=0}^{d-1}\sum_{k=0}^{D-1} \frac{c_{jk}}{\sqrt{d}}\ket{j}_A\ket{k}_B\ket{l}_{A'}\ket{l}_{B'}.
        $$
        Following Eq.~\eqref{Eq:BellStates}, we can define an orthonormal basis of maximally entangled states for the subsystems in $A$ and $A'$,
        $$
            \ket{\phi_{nm}}_{AA'} = \frac{1}{\sqrt{d}}\sum_{r=0}^{d-1} e^{2\pi i r n / d} \ket{r}_{A}\otimes \ket{r \oplus_{d} m}_{A'}.
        $$
        If Alice performs a measurement of this basis on her systems $AA'$ and gets result $n,m$, Bob's post-measurement state (up to normalization) becomes
        \begin{align*}
            \bra{\phi_{nm}}_{AA'} \left(\ket{\varphi_x}_{AB}\ket{\phi_{00}}_{A'B'} \right)
            &=\sum_{j,l,r=0}^{d-1}\sum_{k=0}^{D-1}\frac{c_{jk}}{d}e^{-2\pi i r n / d} \braket{r}{j}_{A}\braket{r \oplus_d m}{l}_{A'}\ket{k}_{B}\ket{l}_{B'}\\
            &=\sum_{r=0}^{d-1}\sum_{k=0}^{D-1}\frac{c_{rk}}{d}e^{-2\pi i r n / d}\ket{k}_{B}\ket{r \oplus_d m}_{B'}.
        \end{align*}
        As the above state is given by a linear combination of orthogonal states, the square of its norm is $1/d^2$.
        Therefore, considering normalization, Bob's post-measurement state is given by
        \begin{align*}
            \sum_{r=0}^{d-1}\sum_{k=0}^{D-1} c_{rk}e^{-2\pi i r n / d}\ket{k}_{B}\ket{r \oplus_d m}_{B'}.
        \end{align*}
        After getting the measurement result (dits $n$ and $m$), Alice sends them to Bob through their classical communication channel. Informed by that, Bob applies the unitary $(\mathbb{1}\otimes U_{nm})$ over his pair of systems $B,B'$, where $U_{nm}$ is given by Eq.~\eqref{Eq:Unm}. He ends up with the state
        \begin{align*}
            \sum_{r,s=0}^{d-1}\sum_{k=0}^{D-1} c_{rk}e^{-2\pi i (r-s) n / d}\braket{s \oplus_d m}{r \oplus_d m}_{B'} \ket{k}_{B}\ket{s}_{B'}
            = \sum_{r=0}^{d-1}\sum_{k=0}^{D-1}c_{rk}\ket{k}_{B}\ket{r}_{B'} .
        \end{align*}

        With this scheme, Alice and Bob carry out a teleportation from $A$ to $B'$. At this point,  to simulate the behavior \eqref{Eq:EA-pm-quantico}, Bob can now apply the same set of measurements $\{M_{b \vert y}\}$ on the state obtained by him after the teleportation process. It is worth noting that the teleportation was done in such a way that the correlations between system $A$ and $B$ are now preserved between systems $B$ and $B'$. In fact, Bob’s final state (up to a choice of basis) is equal to the state $\ket{\varphi_x}$. It is also worth noting that the bases are independent of Alice’s input $x$, so Bob does not need to know Alice’s input $x$ in order to apply the unitary $U_{nm}$.
        
        In the case where $\varrho_{AB}^{x} $ is mixed, let $\varrho_{AB}^{x}  = \sum_{\alpha} \lambda_{\alpha}\ketbra{\varphi_{\alpha}^{x}}{\varphi_{\alpha}^{x}}$ be a decomposition of $\varrho_{AB}^{x} $ as a sum of pure states (spectral decomposition). For each $\alpha$, there are coefficients $\{c_{r,s}^{\alpha}\}$ such that $\ket{\varphi_{\alpha}^{x}} = \sum_{j=0}^{d-1}\sum_{k=0}^{D-1} c_{jk}^{\alpha}\ket{j}_A\ket{k}_B$. Because all the transformations described in the pure state case are linear, the previous steps are also valid for the convex mixture of the states $\ket{\varphi_{\alpha}^{x}}$. So, even when the state $\varrho_{AB}^{x}  = (\mathcal{C}_x \otimes \mathbb{1})(\rho_{AB})$ is mixed, Alice and Bob are able to teleport Alice's part of this state to Bob by sending only a couple of dits and being assisted by the state $\rho_{AB}\otimes \ketbra{\phi_{00}}{\phi_{00}}$.
    \end{proof}

\section{\texorpdfstring{Bounds for $\mathcal{F}_d$}{Bounds for Fd}}\label{sec:Bounds-Fd}

    In Sec. \ref{sec:highdim}, we postulated the following inequality: 
    \begin{equation}\label{eq: Desigualdade F_d_app}
        \mathcal{F}_{d}[\textbf{p}(b|x)] =  2\sum_{i=0}^{d} p(i|i) + \sum_{i=0}^{d}p(d+1|i),
    \end{equation}
    claiming that its maximum value in $C_{d}(d+1,1,d+2)$ is $2d$ and that it can be violated with entanglement assistance of a pair of qubits.
    
    \subsection{\texorpdfstring{Strategy to reach $\mathcal{F}_d = 2d$ in $C_d(d+1,1,d+2)$}{Strategy to reach Fd = 2d in Cd(d+1,1,d+2)}}
        Consider the deterministic strategy where Alice applies the encoder function $\mathcal{E}: [d+1] \to [d]$ on the input received by her, where
        \begin{equation*}
             \mathcal{E}(k) = 
                \begin{cases}
                    k, \ \ \mbox{if} \ 0 \le k \le d-1 \\
                    0, \ \ \mbox{if} \ k=d,
                \end{cases}
        \end{equation*}
        and Bob decodes his received dit via
        \begin{align}
            \mathcal{D} :& \ [d] \to [d+2] \\
                & \ \ k  \  \mapsto \ k \nonumber
        \end{align}
        The behavior thus obtained has elements
        \begin{align}
            q(b|x) = \delta(b, (\mathcal{D} \circ \mathcal{E})(x))
            = \begin{cases}
                        \delta(b,x), \ \ \mbox{if} \ 0 \le x \le d-1 \\
                        \delta(b,0), \ \ \mbox{if } x=d.
                \end{cases}
        \end{align}
        We claim that with this strategy Alice and Bob reach the value of $2d$ for the functional \eqref{eq: Desigualdade F_d_app}. Indeed,
        %
        \begin{align}
            \mathcal{F}_{d}[\textbf{q}(b|x)] = &2\left(\sum_{i=0}^{d-1} \delta(i,i) + \delta(d,0)\right)
            + \sum_{i=0}^{d-1} \delta(d+1,i) +  \delta(d+1,0)
            = 2d.
        \end{align}

    \subsection{Optimality of the strategy}
        We will now show that the maximum of $\mathcal{F}_{d}$ in $C_{d}(d+1,1,d+2)$ is in fact equal to $2d$. Because $C_{d}(d+1,1,d+2)$ is a polytope and $\mathcal{F}_d$ is a linear functional of the behaviors, the maximum occurs at the extreme points of this polytope, which are given by the deterministic encoding/decoding strategies involving one dit of communication. Thus, it suffices to show that the maximum of $\mathcal{F}_{d}$ over the deterministic strategies is in fact $2d$.
        
        Let $q(b|x)$ be a deterministic point of $C_{d}(d+1,1,d+2)$. Then, there is an $\mathcal{E}: [d+1] \to [d] $ and a $\mathcal{D}: [d] \to [d+2]$ such that $q(b|x) = \delta\left[ b, (\mathcal{D} \circ \mathcal{E})(x) \right]$.  Since the domain of $\mathcal{D}$ has cardinality $d$, it follows that the cardinality of the image of $\mathcal{D}$ is smaller or equal to $d$, \textit{i.e}, $|\text{Im}(\mathcal{D})| \le d$. On the other hand, $\text{Im}(\mathcal{D} \circ \mathcal{E}) \subseteq \text{Im}(\mathcal{D})$. Therefore, it follows that $|\text{Im}(\mathcal{D} \circ \mathcal{E})| \le d$. 
        
        At this point, we divide the problem into two situations. In the first one, we will assume that $d+1 \in \text{Im}(\mathcal{D} \circ \mathcal{E})$. As we also know that $|\text{Im}(\mathcal{D} \circ \mathcal{E})|\le d$ and $\text{Im}(\mathcal{D} \circ \mathcal{E}) \subset [d+2] $, it follows that there are $k,l \in [d+1]$ such that $k,l \notin \text{Im}(\mathcal{D} \circ \mathcal{E})$. Remembering that $q(b|x) = \delta\left[ b, (\mathcal{D} \circ \mathcal{E})(x))\right]$, then it follows that $q(k|k) = 0 = q(l|l)$, so:
        \begin{align}\label{eq: F_d caso 1}
            \mathcal{F}_{d}\left[ \textbf{q}(b|x))\right] = &2q(k|k) + 2q(l|l) + q(d+1|k) + q(d+1|l)
            + \sum_{\substack{i=0 \\ i \neq l,k}}^{d} \left[ 2q(i|i) + q(d+1|i) \right].
        \end{align}
        But, from normalization, we have the following upper bounds for the probabilities appearing in \eqref{eq: F_d caso 1}:
        \begin{subequations}
        \begin{align}
            &q(d+1|k) \le 1, \label{eq: q(d+1|k)} \\
            &q(d+1|l) \le 1, \label{eq: q(d+1|l)} \\
            &2q(i|i) + q(d+1|i) \le 2(q(i|i) + q(d+1|i)) \le 2. \label{eq: soma q}
        \end{align}
        \end{subequations}
        Replacing them in \eqref{eq: F_d caso 1} and using that $q(k|k) = 0 = q(l|l)$, we reach
        \begin{align}
            \mathcal{F}_{d}\left[ \textbf{q}(b|x) \right] \le 2 + 2(d - 1) = 2d.
        \end{align}
        
        Suppose now that $d+1 \notin \text{Im}(\mathcal{D} \circ \mathcal{E})$ which implies that $\text{Im}(\mathcal{D} \circ \mathcal{E}) \subset [d+1]$ and $q(d+1|i) = \delta\left[ d+1, (\mathcal{D} \circ \mathcal{E})(i) \right] = 0$ for all $i \in [d+1]$. Thus,
        \begin{align*}
            \mathcal{F}_d\left[ \textbf{q}(b|x) \right] &=  2\sum_{i=0}^{d}q(i|i).
        \end{align*}
        On the other hand, as $|\text{Im}(\mathcal{D} \circ \mathcal{E})| \le d$, it follows that there is $k \in [d+1]$ such that $k \notin \text{Im}(\mathcal{D} \circ \mathcal{E})$, which implies $q(k|k)=0$. Then, using that $q(i|i) \le 1$ for all $i \in [d+1]$,
        \begin{align*}
            \mathcal{F}_d \left[ \textbf{q}(b|x))\right] &= 2\sum_{\substack{i=0 \\ i \neq k}}^{d}q(i|i) \leq 2d .
        \end{align*}

    \subsection{\texorpdfstring{Entanglement-assisted violation of $\mathcal{F}_d \leq 2d$}{Entanglement-assisted violation of Fd <= 2d}}
    
        It is always possible to obtain a value for $\mathcal{F}_d$ greater than $2d$ when considering the entanglement-assisted behaviors in $C_{d}^{2}$. For this, we will use the behavior $ \mathbf{p^\prime}(b|x) \in C_2^2 (3, 1, 4)$ obtained in \cite{repo} --- namely, the one such that $\mathcal{F}_{2}[\mathbf{p^\prime}(b|x)] = 4.155$,--- to build a $\textbf{p}^{*}(b|x)$ composed of
        \begin{equation}\label{eq: def p^*}
            p^{*}(b|x)= \begin{cases}
                            p'(b|x), \ \ \mbox{if } 0 \le x \le 2 \mbox{ and } 0 \le b \le 3\\
                            0, \ \ \ \ \ \ \ \ \ \mbox{if } 0 \le x \le 2 \mbox{ and }  b > 3\\
                            \delta(b,x), \ \ \mbox{if } 3 \le x \le d.
                        \end{cases}
        \end{equation}
        We can see that $\textbf{p}^*$ belongs to $C_{d}^{2}$. In fact, since Alice can send a $d$it to Bob, she uses 2 symbols from this alphabet to simulate the $\textbf{p}'(b|x)$ part of the behavior, while the remaining $d-2$ symbols are used to simulate the other $\delta(b,x)$ part of the behavior $\textbf{p}^{*}(b|x)$.
        
        Substituting into $\mathcal{F}_{d}$,
        %
        \begin{equation*}
            \mathcal{F}_{d}\left[ \textbf{p}^{*}(b|x))\right]
            = \mathcal{F}_{2}[\mathbf{p^\prime}(b|x)] 
            + 2\sum_{i=3}^{d}p^*(i|i) + \sum_{i=3}^{d}p^*(d+1|i).
        \end{equation*}
        %
        But, from Eq.~\eqref{eq: def p^*},
        \begin{equation*}
            2\sum_{i=3}^{d}p^*(i|i) = 2\sum_{i=3}^{d} \delta(i,i) = 2(d-2),
        \end{equation*}
        and
        \begin{equation*}
            \sum_{i=3}^{d}p^*(d+1|i) = \sum_{i=3}^{d} \delta(d+1,i) = 0.
        \end{equation*}
        Therefore, 
        \begin{align}
            \mathcal{F}_{d}\left[ \textbf{p}^{*}(b|x))\right] = 4.155 + 2(d-2) > 2d.
        \end{align}
        Since the maximum of $\mathcal{F}_d$ in $C_{d}$ is equal to $2d$, it follows that $p^{*}(b|x) \notin C_{d}$ .
\end{document}